\documentclass[twoside,a4paper]{amsart}
\usepackage{graphicx}

\PassOptionsToPackage{pdfauthor={Vladimir V. Kisil},%
    pdftitle={Two-Dimensional Conformal Models of Space-Time and Their Compactification},%
    pdfsubject={Mathematical physics},%
    backref=page,%
  pdfkeywords={symmetries}
}{hyperref}
\IfFileExists{eulervm.sty}{\usepackage{eulervm}
}{}
\newcommand{\cycle}[3][]{{#1 C^{#2}_{#3}}}

\newcommand{\zcycle}[3][]{#1 Z^{#2}_{#3}}
\newcommand{\DSpace}[2]{\ensuremath{ { \dot{\mathbb{#1}}^{#2}} }}
\newcommand{\TSpace}[2]{\ensuremath{ { \widetilde{\mathbb{#1}}^{#2}} }}
\providecommand{\bs}{\breve{\sigma}}
\providecommand{\rmc}{\mathrm{\breve\i}}

\usepackage[printedin]{myamsart}
\ppnum{LEEDS--MATH--PURE--2006--17
}{arXiv: \href{http://arXiv.org/abs/math-ph/0611053}{math-ph/0611053}}
{2006}

\input{mydef}

\begin{document}
\title[2D Model of Conformal Space-Time]{Two-Dimensional Conformal Models of Space-Time and Their Compactification}

\author[Vladimir V. Kisil]%
{\href{http://maths.leeds.ac.uk/~kisilv/}{Vladimir V. Kisil}}
\thanks{On  leave from Odessa University.}

\address{%
School of Mathematics\\
University of Leeds\\
Leeds LS2\,9JT\\
UK
}

\email{\href{mailto:kisilv@maths.leeds.ac.uk}{kisilv@maths.leeds.ac.uk}}

\urladdr{\href{http://maths.leeds.ac.uk/~kisilv/}%
{http://maths.leeds.ac.uk/\~{}kisilv/}}

\maketitle

\begin{abstract}
  We study geometry of two-dimensional models of conformal space-time
  based on the group of M\"obius transformation. The natural geometric
  invariants, called cycles, are used to linearise M\"obius action.
  Conformal completion of the space-time is achieved through an
  addition of a zero-radius cycle at infinity. We pay an attention to
  the natural condition of non-reversibility of time arrow in order to
  get a correct compactification in the hyperbolic case.
\end{abstract}

\section{Introduction}
\label{sec:introduction}

The ideas of symmetries and invariants were introduced by Galileo
centuries ago but still are central to many branches of
theoretical physics. In a mathematical version they are known as \href{http://en.wikipedia.org/wiki/Erlangen_program}{Erlangen
program} delivered by
\href{http://www-groups.dcs.st-and.ac.uk/~history/Mathematicians/Klein.html}{F.~Klein}
and influenced by
\href{http://turnbull.mcs.st-and.ac.uk/~history/Mathematicians/Lie.html}{S.~Lie},
which is mistakenly limited to geometry
often~\cite{Kisil02c,Kisil06a}.

Study of space-time geometry based on the
group of conformal maps was very fruitful, see for
example~\cite{Segal76}. It is worth o consider a two-dimensional
conformal space-time~\cite{HerranzSantander02b,HerranzSantander02a} as
it illustrates many general features.

In this paper we briefly overview an
approach~\cite{Cnops02a,Kisil05a,Kisil06a} to conformal space-time
geometry associated with the M\"obius action of \(\SL\) group---the
group of \(2\times 2\) matrices of real entries with the unit
determinant. We emphasise natural conformal invariants---called
cycles~\cite{Yaglom79}---and their r\^ole in conformal completion of
the space-time. The usage of hypercomplex (dual and double) numbers
along with complex ones gives a unified description for all model. We
also stress some natural physical requirements omitted
in~\cite{HerranzSantander02b}.  Although all mentioned bits appeared
in literature in different places before their gathering under one
roof seems to be new and quite fruitful. Moreover it opens a
straightforward possibility for a multidimensional generalisation, which
is outlined at the end of paper. 

\section[Moebius transformations and cycles]{M\"obius transformations and cycles}
\label{sec:make-guess-three}

A left  action of \(\SL\) group on a real line is given by linear-fractional or M\"obius maps:
\begin{equation}
  \label{eq:moebius}
  g: x\mapsto g\cdot x=\frac{ax+b}{cx+d}, \text{ where } 
  g=  \begin{pmatrix}
    a&b\\c&d
  \end{pmatrix}\in\SL,\  x\in\Space{R}{}.
\end{equation}
This action makes
sense also as a map of \emph{complex},  \emph{double} and \emph{dual}
numbers \cite{Kisil06a}, \cite[Suppl.~C]{Yaglom79}, which have the form \(z=x+\rmi y\),
with different values of the imaginary unit square: \(\rmi^2=-1\), \(0\) or
\(1\) correspondingly. Although the arithmetic of dual and
double numbers are different from the complex ones, e.g. they have
divisors of zero, there are many properties in common.

Three possible values \(-1\), \(0\) and \(1\) of \(\sigma:=\rmi^2\)
will be refereed here as \emph{elliptic}, \emph{parabolic} and
\emph{hyperbolic} cases respectively. All three together will be
abbreviated as EPH. One of the reason is that circles, parabolas and
(equilateral) hyperbolas are invariant under the corresponding
M\"obius transformation.

The common name \emph{cycle}~\cite{Yaglom79} is used to denote the
corresponding conic sections (as well as straight lines as their
limits) in the respective EPH case. More precisely a generic cycle is
the set of points \((u,v)\in\Space{R}{2}\) defined for all values of
\(\sigma\) by the equation
\begin{equation}
  \label{eq:cycle-eq}
  k(u^2-\sigma v^2)-2lu-2nv+m=0.
\end{equation}
We denote this space by \(\Space{R}{\sigma}\) for a generic
\(\sigma\), specialised EPH cases will be written as \(\Space{R}{e}\),
\(\Space{R}{p}\), \(\Space{R}{h}\).

A common method is to declare objects under investigations (cycles in
our case, functions in functional analysis, etc.) to be simply points
of some bigger space.  Equation~\eqref{eq:cycle-eq} (and the
corresponding cycle) is defined by a point \((k, l, n, m)\) from a
projective space \(\Space{P}{4}\), since for a scaling factor
\(\lambda \neq 0\) the point \((\lambda k, \lambda l, \lambda n,
\lambda m)\) defines the same equation~\eqref{eq:cycle-eq}. We call
\(\Space{P}{4}\) the \emph{cycles space} and refer to the initial
\(\Space{R}{2}\) as the \emph{points space}.

In order to get a connection with M\"obius action~\eqref{eq:moebius}
we arrange numbers \((k, l, n, m)\) into the \(2\times 2\) matrix
\cite{Cnops02a,Kisil05a,Kisil06a} 
\begin{equation}
  \label{eq:FSCc-matrix}
  C_{\bs}^s=\begin{pmatrix}
    l+\rmc s n&-m\\k&-l+\rmc s n
  \end{pmatrix}, 
\end{equation}
with a new imaginary unit \(\rmc\) and an additional parameter \(s\)
usually equal to \(\pm 1\). The values of \(\bs:=\rmc^2\) is \(-1\),
\(0\) or \(1\) independently from the value of \(\sigma\).  The
matrix~\eqref{eq:FSCc-matrix} is the cornerstone of (extended)
Fillmore--Springer--Cnops construction
(FSCc)~\cite{Cnops02a,Kirillov06,Kisil05a,Kisil06a}. 

The FSCc is significant for conformal maps because it \emph{intertwines}
M\"obius action~\eqref{eq:moebius} on cycles with
a linear map by matrix similarity:
\begin{equation}
  \label{eq:cycle-similarity}
  \tilde{C}_{\bs}^s= gC_{\bs}^sg^{-1}.
\end{equation}

\section{Invariants of cycles: algebraic and geometric}
\label{sec:invar-algebr-geom}

For \(2\times 2\) matrices (and thus cycles) there are only two
essentially different invariants under
similarity~\eqref{eq:cycle-similarity} (and thus under M\"obius
action~\eqref{eq:moebius}): the \emph{trace} and the
\emph{determinant}. However due to projective nature of the
cycle space \(\Space{P}{4}\) the a non-zero values of trace or
determinant are relevant only if cycles are suitable normalised.

For example, if \(k\neq0\) we may
normalise the quadruple to
\((1,\frac{l}{k},\frac{n}{k},\frac{m}{k})\) with highlighted cycle's
centre. Moreover in this case \(\det \cycle{s}{\bs}\) is equal to
the square of cycle's radius~\cite{Kisil05a,Kisil05b}.
Another normalisation \(\det \cycle{s}{\bs}=1\) is used
in~\cite{Kirillov06} to get a nice condition for touching circles.

We still get important characterisation even with non-normalised
cycles, e.g., invariant classes (for different \(\bs\)) of
cycles are defined by the condition \(\det C_{\bs}^s=0\). Such a
class is parametrises only by two real number and as such is easily
attached to certain point of \(\Space{R}{2}\). For example, the
cycle \(C_{\bs}^s\) with \(\det C_{\bs}^s=0\), \(\bs=-1\) drawn
elliptically represent just a point \((\frac{l}{k},\frac{n}{k})\),
i.e. (elliptic) \emph{zero-radius circle}.  The same condition with
\(\bs=1\) in hyperbolic drawing produces a null-cone originated at
point \((\frac{l}{k},\frac{n}{k})\):
\begin{displaymath}
  (u-\frac{l}{k})^2-(v-\frac{n}{k})^2=0,
\end{displaymath}
i.e. a zero-radius cycle in hyperbolic metric. Note that in all cases
the zero-radius cycle consists of non-invertible points (divisor of
zero) in \(\Space{R}{\sigma}\). We denote a zero-radius cycle centred
at \((u,v)\) by \(\zcycle{}{(u,v)}\).

\begin{rem}
  In general for every notion there is nine possibilities: three EPH
  cases (three values of \(\bs\)) in the cycle space times three EPH
  realisations (three values of \(\sigma\)) in the point space.

  We should clarify the relation between our nine cases and nine
  Cayley-Klein
  geometries~\cite{HerranzSantander02a,HerranzSantander02b,Yaglom79}.
  Our parameter \(\sigma\) describes the signature of the point space:
  (which is \((-1,\sigma)\)) and thus \(\sigma\) coincides with parameter \(\kappa_2\)
  from~\cite{HerranzSantander02a,HerranzSantander02b}. However our
  parameter \(\bs\) is \textbf{different} from \(\kappa_1\) used
  in~\cite{HerranzSantander02a,HerranzSantander02b}. The later
  describes constant curvature whereas all our spaces are ``flat''
  (their curvature is zero). Our parameter \(\bs\) describes metric of the
  cycle spaces \(\Space{P}{4}\):  if it is different from \(\sigma\) then this
  measure ``non-locality'' of our geometry. This notion cannot be
  embedded in any way into the Riemann geometry since it is ``local''
  from the definition.
\end{rem}

The most expected relation between cycles is based on the following
M\"obius invariant ``inner product'' build from a trace of product of
two cycles as matrices (cf. with GNS construction to make a Hilbert
space out of \(C^*\)-algebra~\cite{Arveson76}):
\begin{equation}
  \label{eq:inner-prod}
  \scalar{C_{\bs}^s}{\tilde{C}_{\bs}^s}= \tr (C_{\bs}^s\tilde{C}_{\bs}^s)
\end{equation}
The next standard move is to define  \emph{\(\bs\)-orthogonality} of 
two cycles as vanishing of their inner product:
\begin{equation}
  \label{eq:ortho}
  \scalar{C_{\bs}^s}{\tilde{C}_{\bs}^s}=0. 
\end{equation}

For the case of \(\bs \sigma=1\), i.e. when geometries of the cycles
and points spaces are both either elliptic or hyperbolic, such an
orthogonality is the standard one, defined in terms of 
angles between tangent lines in the intersection points of two
cycles. However in the remaining seven (\(=9-2\)) cases the
innocent-looking definition~\eqref{eq:ortho} brings unexpected
relations between cycles~\cite{Kisil05a}.

One can easy check the following orthogonality properties of the
zero-radius cycles defined in the previous section~\cite{Kisil05a}:
\begin{enumerate}
\item Since \(\scalar{C_{\bs}^s}{{C}_{\bs}^s}=\det {C}_{\bs}^s\)
  zero-radius cycles are  self-orthogonal (isotropic) ones.
\item \label{it:ortho-incidence}
  A cycle \(\cycle{s}{\bs}\) is \(\sigma\)-orthogonal to a zero-radius
  cycle \(\zcycle{s}{\bs}\) if and only if \(\cycle{s}{\bs}\) passes
  through the \(\sigma\)-centre of \(\zcycle{s}{\bs}\).
\end{enumerate}

\section{Conformal compactification by zero-radius cycles}
\label{sec:conf-comp}

\begin{figure}[tbp]
  \centering
  (a)\includegraphics{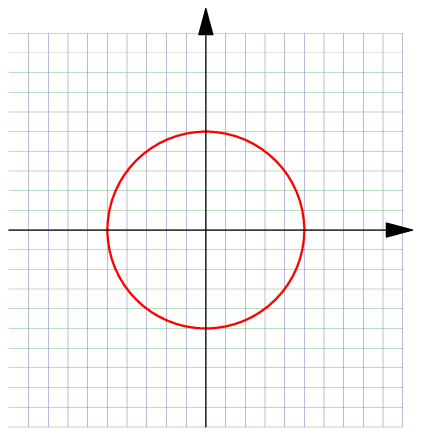}\qquad
  (b)\includegraphics{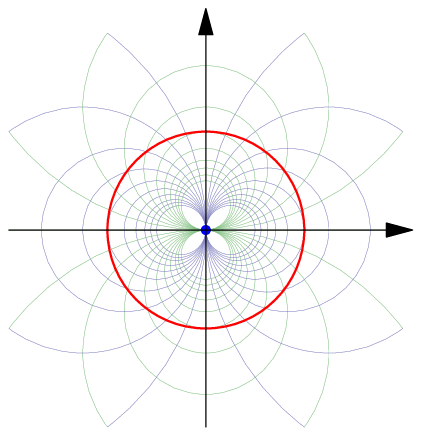}\\[5pt]
  (c)\includegraphics{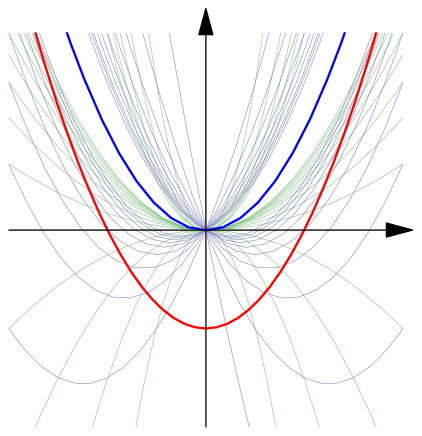}\qquad
  (d)\includegraphics{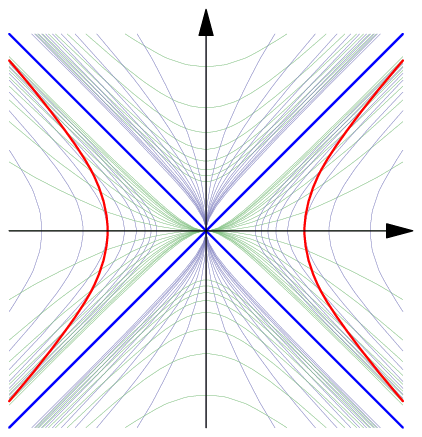}
  \caption[Three types of inversions of the rectangular grid]{Three
    types of inversions of the rectangular grid. The initial
    rectangular grid (a) is inverted elliptically in the unit circle
    (shown in red) on (b), parabolically on (c) and hyperbolically on
    (d). The blue cycle (collapsed to a point at the origin on (a))
    represent the image of the cycle at infinity under inversion.}  
  \label{fig:inversions}
\end{figure}

Likewise an association of points of \(\Space{P}{4}\) with
cycles~\eqref{eq:cycle-eq} in \(\Space{R}{\sigma}\) overlooked
exceptional values. For example, the image of
\((0,0,0,1)\in\Space{P}{4}\), which is corresponds to the equation
\(1=0\), is not a valid conic section in \(\Space{R}{\sigma}\). We
also did not investigate yet accurately singular points of the
M\"obius map~\eqref{eq:moebius}. It turns out that both questions are
connected.

One of the standard approaches~\cite[\S~1]{Olver99} to deal with
singularities of the M\"obius map is to consider projective coordinates
on the plane. Since we have already a projective space of cycles, we
may use it as a model for compactification which is even more
appropriate. The identification of points with zero-radius cycles,
discussed above, plays an important r\^ole here.

More specifically we represent the  \emph{zero-radius cycle at
  infinity} by the point  \((0,0,0,1)\in\Space{P}{4}\) denoted it by
\(\zcycle{}{\infty}\). The following results on \(\zcycle{}{\infty}\)
can be found in~\cite{Kisil05a}:
\begin{enumerate}
\item \(\zcycle{}{\infty}\) is the image of the zero-radius cycle
  \(\zcycle{}{(0,0)}=(1,0,0,0)\) at the origin under reflection
  (inversion) into the unit cycle \((1, 0,0,-1)\), see blue cycles
  in Fig.~\ref{fig:inversions}(b)-(d).
\item The following statements are equivalent
  \begin{enumerate}
  \item A point \((u,v)\in\Space{R}{\sigma}\) belongs to the
    zero-radius cycle \(\zcycle{}{(0,0)}\) centred at the origin;
  \item The zero-radius cycle \(\zcycle{}{(u,v)}\) is \(\sigma\)-orthogonal to 
    zero-radius cycle \(\zcycle{}{(0,0)}\);
  \item The inversion \(z\mapsto \frac{1}{z}\) in the unit cycle is
    singular in the point \((u,v)\);
  \item The image of \(\zcycle{}{(u,v)}\) under inversion in the unit
    cycle is orthogonal to  \(\zcycle{}{\infty}\). 
  \end{enumerate}
  If any from the above is true we also say that image of \((u,v)\) under
  inversion in the unit cycle belongs to zero-radius cycle at infinity.
\end{enumerate}
  
\begin{figure}[htbp]
  \centering
  \includegraphics
{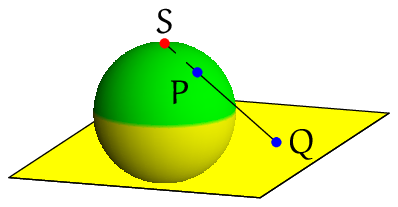} \hfill
  \includegraphics
{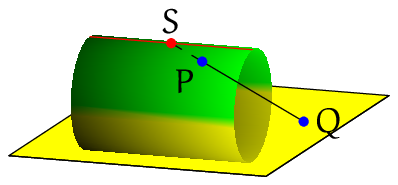} \hfill
  \includegraphics
{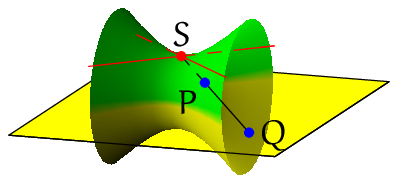}\\
(a)\hfill(b)\hfill(c)\hfill
  \caption[Compactification and stereographic
  projections]{Compactification of \(\Space{R}{\sigma}\) and
    stereographic projections.} 
  \label{fig:compactifications}
\end{figure}

Now we are going to discuss conformal compactification of spaces
\(\Space{R}{\sigma}\) which makes action of M\"obius
map~\eqref{eq:moebius} on it non-singular at any point.  Our
description will be based on the concept of zero-radius cycle.

It is very well known that in the elliptic case the conformal
compactification is done by addition to \(\Space{R}{e}\) a point
\(\infty\) at infinity, which is the elliptic zero-radius cycle.
However in the parabolic and hyperbolic cases the singularity of the
M\"obius transform is not localised in a single point---the
denominator is a zero divisor for the whole zero-radius cycle.  Thus
in each EPH case the correct compactification is made by the union
\(\Space{R}{\sigma}\cup\zcycle{}{\infty}\).

 It is common to identify the compactification \(\DSpace{R}{e}\) of the
 space \(\Space{R}{e}\) with a Riemann sphere.
 This model can be visualised by the stereographic
 projection, see~\cite[\S~18.1.4]{BergerII} and
 Fig.~\ref{fig:compactifications}(a). A similar model can be 
 provided for the parabolic and hyperbolic spaces as
 well, see~\cite{HerranzSantander02b} and
 Fig.~\ref{fig:compactifications}(b),(c). Indeed the space 
 \(\Space{R}{\sigma}\) can be identified with a corresponding surface
 of the constant curvature: the sphere (\(\sigma=-1\)), the cylinder
 (\(\sigma=0\)), or the one-sheet hyperboloid (\(\sigma=1\)). The map
 of a surface to \(\Space{R}{\sigma}\) is given by the polar
 projection, see~\cite[Fig.~1]{HerranzSantander02b} and
 Fig.~\ref{fig:compactifications}(a)-(c). These
 surfaces provide ``compact'' model of the corresponding
 \(\Space{R}{\sigma}\) in the sense that M\"obius transformations
 which are lifted from \(\Space{R}{\sigma}\) by the projection are
 not singular on these surfaces. 

 However the hyperbolic case has its own caveats which may be easily
 oversight, cf.~\cite{HerranzSantander02b}. A
 compactification of the hyperbolic space \(\Space{R}{h}\) by a light
 cone (which the hyperbolic zero-radius cycle) at infinity will indeed
 produce a closed M\"obius invariant object. However it will not be
 satisfactory for some other reasons, physical in particular, which
 explained in the next subsection. 

\section{Non-invariance of the upper half-plane}
\label{sec:invar-upper-half}
The important difference between the hyperbolic case and the two
others is that in the elliptic and parabolic cases the upper halfplane
in \(\Space{R}{\sigma}\) is preserved by M\"obius transformations from
\(\SL\). However in the hyperbolic case any point \((u,v)\) with
\(v>0\) can be mapped to an arbitrary point \((u',v')\) with \(v'\neq
0\).

The lack of invariance in the hyperbolic case has many important
consequences in seemingly different areas, for example:
\begin{figure}[tbp]
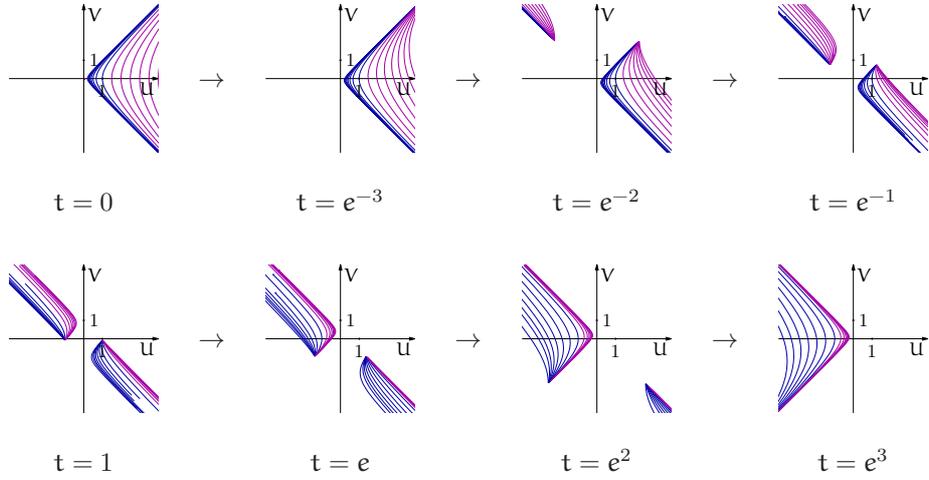

  \parbox[t]{.2\textwidth}{
    \begin{center}
      \includegraphics[scale=.70]{parabolic0.20}\\
      \(t=0\)
    \end{center}
  } \hfill\raisebox{-1.2cm}{\(\to\)} \hfill
  \parbox[t]{.2\textwidth}{
    \begin{center}
      \includegraphics[scale=.70]{parabolic0.21}\\
      \(t=e^{-3}\)
    \end{center}
  }   \hfill \raisebox{-1.2cm}{\(\to\)} \hfill
  \parbox[t]{.2\textwidth}{
    \begin{center}
      \includegraphics[scale=.70]{parabolic0.22}\\
      \(t=e^{-2}\)
    \end{center}
  } \hfill \raisebox{-1.2cm}{\(\to\)} \hfill
    \parbox[t]{.2\textwidth}{
    \begin{center}
      \includegraphics[scale=.70]{parabolic0.23}\\
      \(t=e^{-1}\)
    \end{center}
  }\\[5mm]
  \parbox[t]{.2\textwidth}{
    \begin{center}
      \includegraphics[scale=.70]{parabolic0.24}\\
      \(t=1\)
    \end{center}
  }  \hfill \raisebox{-1.2cm}{\(\to\)} \hfill
  \parbox[t]{.2\textwidth}{
    \begin{center}
      \includegraphics[scale=.70]{parabolic0.25}\\
      \(t=e\)
    \end{center}
  } \hfill \raisebox{-1.2cm}{\(\to\)} \hfill
  \parbox[t]{.2\textwidth}{
    \begin{center}
      \includegraphics[scale=.70]{parabolic0.26}\\
      \(t=e^2\)
    \end{center}
  } \hfill \raisebox{-1.2cm}{\(\to\)} \hfill
  \parbox[t]{.2\textwidth}{
    \begin{center}
      \includegraphics[scale=.70]{parabolic0.27}\\
      \(t=e^3\)
    \end{center}
  }
  \caption[Continuous transformation from future
    to the past]{Eight frames from a continuous transformation from future
    to the past parts   of the light cone.}
  \label{fig:future-to-past}
\end{figure}
\begin{description}
\item[\textbf{Geometry}] \(\Space{R}{h}\) is not split by the real
  axis into two disjoint pieces: there is a continuous path (through
  the light cone at infinity) from the upper half-plane to the lower which
  does not cross the real axis (see \(\sin\)-like curve joined two sheets of
  the hyperbola in Fig.~\ref{fig:hyp-upper-half-plane}(a)).
\item[\textbf{Physics}] There is no M\"obius invariant way to separate
  ``past'' and ``future'' parts of the light cone, see
  \cite[\S~III.4]{Segal76} for a detailed discussion of this effect
  and implications in the four-dimensional Minkowski space-time. More
  precisely it it may be stated as follows: there is a continuous
  family of M\"obius transformations reversing the arrow of time. For
  example, the family of matrices \(
  \begin{pmatrix}
    1&-t\rmi\\t\rmi&1
  \end{pmatrix}\), \(t\in [0,\infty)\) provides such a transformation.
  Fig.~\ref{fig:future-to-past} illustrates this by corresponding
  images for eight subsequent values of \(t\).
\item[\textbf{Analysis}] There is no a possibility to split
  \(\FSpace{L}{2}(\Space{R}{})\) space of function into a direct sum
  of the Hardy type space of functions having an analytic extension into
  the upper half-plane and its non-trivial complement, i.e. any function
  from \(\FSpace{L}{2}(\Space{R}{})\) has an ``analytic extension''
  into the upper half-plane in the sense of hyperbolic function
  theory,  see~\cite{Kisil97c}.
\end{description}
\begin{figure}[htbp]
  \centering
     (a)\includegraphics[scale=.75]{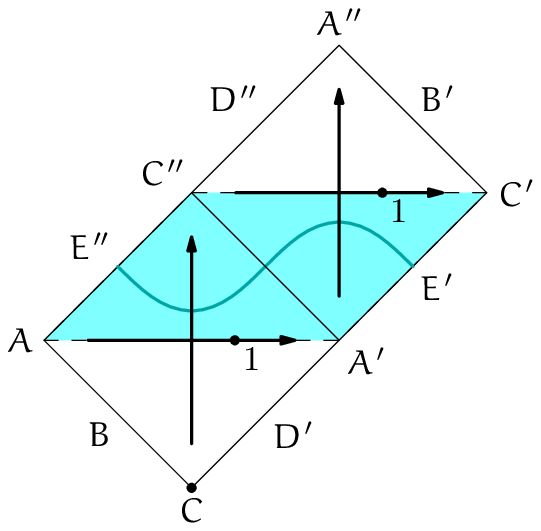} \hspace{.1\textwidth}
    (b)\includegraphics[scale=.75
]{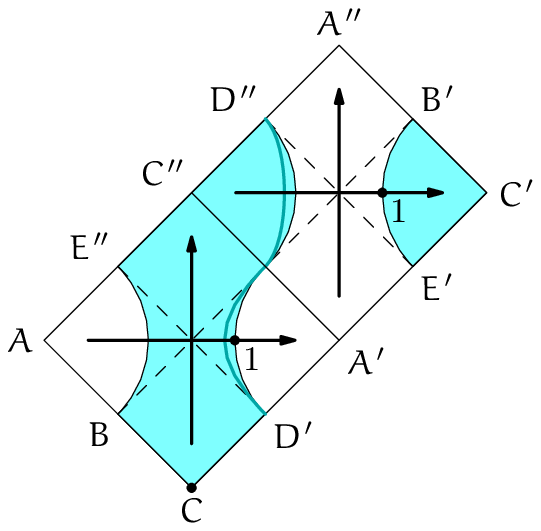}
  \caption[Hyperbolic objects in the double cover]{Hyperbolic objects
    in the double cover of \(\Space{R}{h}\):\\ 
  (a) the ``upper'' half-plane;\qquad (b) the unit circle.}
  \label{fig:hyp-upper-half-plane}
\end{figure} 
All the above problems can be resolved in the following
way~\cite[\S~A.3]{Kisil97c}.  We take two copies \(\Space[+]{R}{h}\)
and \(\Space[-]{R}{h}\) of \( \Space{R}{h} \), depicted by the squares
\(ACA'C''\) and \(A'C'A''C''\) in Fig.~\ref{fig:hyp-upper-half-plane}
correspondingly. The boundaries of these squares are light cones at
infinity and we glue \(\Space[+]{R}{h}\) and \(\Space[-]{R}{h}\) in
such a way that the construction is invariant under the natural action
of the M\"obius transformation.  That is achieved if the letters
\(A\), \(B\), \(C\), \(D\), \(E\) in
Fig.~\ref{fig:hyp-upper-half-plane} are identified regardless of the
number of primes attached to them. 
The corresponding model through a stereographic projection is presented
on Fig.~\ref{fig:compact-2}, compare with Fig.~\ref{fig:compactifications}(c).
\begin{figure}[htbp]
  \centering
  \includegraphics
{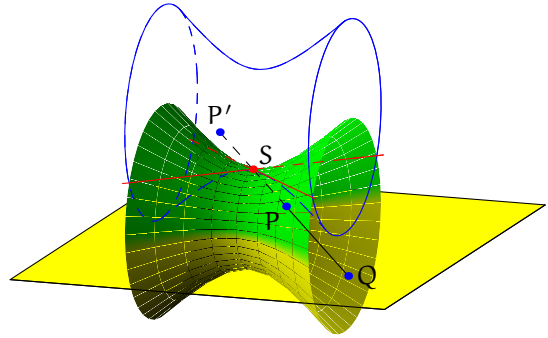}
  \caption[Double cover of the hyperbolic space]{Double cover of the
    hyperbolic space, cf. Fig.~\ref{fig:compactifications}(c). The
  second hyperboloid is shown as a  blue skeleton. It is attached to the
  first one along the light cone at infinity, which is represented by
  two red lines.}  
  \label{fig:compact-2}
\end{figure}

This aggregate denoted by \(\TSpace{R}{h}\) is a two-fold cover of
\(\Space{R}{h}\). The hyperbolic ``upper'' half-plane in
\(\TSpace{R}{h}\) consists of the upper halfplane in
\(\Space[+]{R}{h}\) and the lower one in \(\Space[-]{R}{h}\).  A
similar conformally invariant infinite-fold cover of the Minkowski
space-time was constructed in~\cite[\S~III.4]{Segal76} in connection
with the red shift problem in extragalactic astronomy. A suitable
modification of our double cover of two-dimensional space can be
designed for the conformal version of the four-dimens\-io\-nal Minkowski
space-time.  

The similar construction should be also applied to the conformal
version of the hyperbolic unit disk~\cite{Kisil05a}. It is the image of
the hyperbolic upper-half plane under the Cayley transform defined by
the matrix \(
\begin{pmatrix}
  1&-\rmi\\\rmi&1
\end{pmatrix}\).
 We define it in \( \TSpace{R}{h} \) as follows:
\begin{eqnarray*}
  \TSpace{D}{}&=&\{(u +\rmi v) \such u^2-v^2>-1,\ u\in \Space[+]{R}{h} \}\\
  &&{}\cup \{(u+\rmi v) \such u^2-v^2<-1,\ u\in \Space[-]{R}{h} \}.
\end{eqnarray*}
It can be shown that \( \TSpace{D}{}\) is conformally invariant and 
has a boundary \(\TSpace{T}{}\)---two copies of the unit circles in 
\(\Space[+]{R}{h}\) and \(\Space[-]{R}{h}\). We call \(\TSpace{T}{}\) 
the \emph{(conformal) unit circle} in \(\Space{R}{h}\). 
Fig.~\ref{fig:hyp-upper-half-plane}(b) illustrates
the geometry of the conformal unit disk in \(\TSpace{R}{h}\) in
comparison with the ``upper'' half-plane.

\section{Conclusions and discussions}
\label{sec:conclusion}

This paper discusses conformal geometry of elliptic, parabolic and
hyperbolic two dimensional space-time in term of complex, dual and double
numbers and associated M\"obius transformations. We identify cycles
(i.e. conics of the matching type) as natural invariant objects of the
conformal geometries and associate zero-radius cycles with points of
the space-time. A conformal compactification of the space-time is
uniformly achieved by addition the zero-radius cycle at infinity. Two
other methods (group-theoretical approach and stereographic
projection) were used
in~\cite{HerranzSantander02b,HerranzSantander02a}. 

However in the hyperbolic case the compactification requires more
consideration. In order to forbid a conformal map which revert
time-arrow on the space-time we need to consider (at least) a double
cover of the space-time. Such a model still can be realised through
a suitable stereographic projection.

It is possible that cycles have a deeper physical meaning along
with the geometrical one. It may worth to explore this side of result
from~\cite{Kisil05a}. 

To conclude the discussion we mention a possibility of a higher
dimensions generalisation.  Using Clifford algebras one can
define~\cite{Cnops02a,Porteous95} the Vahlen--Liouville group of
\(2\times 2\) matrices which transform \(\Space{R}{pq}\) mapping
(pseudo)-spheres into (pseudo)-spheres by M\"obius transformations,
where \(\Space{R}{pq}\) is (pseudo)-Euclidean space of signature
\((p,q)\). For the Minkowski space-time the similar construction can
use quaternions instead of the corresponding Clifford algebra. In
higher dimensions our parameter \(\sigma\) will be replaced by the
corresponding signature of the bilinear form defining the metric.
However degenerate bilinear forms seem not be considered in this
context so far.

\textbf{Acknowledgement:} I am grateful to the anonymous referee for
valuable suggestions which help to improve this paper.
\small
\bibliographystyle{plain}
\bibliography{arare,abbrevmr,akisil,aclifford,analyse,aphysics}
\end{document}